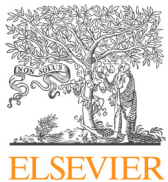

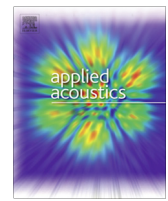

# Auditory distraction in open-plan office environments: The effect of multi-talker acoustics

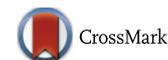

Manuj Yadav *, Jungsoo Kim, Densil Cabrera, Richard de Dear

*School of Architecture, Design and Planning, The University of Sydney, Australia*

## ARTICLE INFO

## ABSTRACT

Within the soundscapes of open-plan offices, irrelevant speech has consistently been reported as the most distracting, and causing performance decrements for workers. Notwithstanding this generalization, the 'babble' created by multiple simultaneously active talkers can sometimes provide beneficial sound masking, but due to spatial release from masking (SRM), speech may still be sufficiently intelligible up to a certain number of talkers (estimated to be about four). This was explored within a highly-realistic office simulation, where the cognitive performance, and subjective distraction of participants were tested. The experimental design was a 4 × 2 factorial (4 talker numbers, 2 levels of broadband sound masking, as the factors). The results indicated that within lower sound pressure level (SPL) of broadband sound masking, multi-talker sound environments degraded cognitive tasks performance more than those with a single talker, suggesting SRM effects. For higher SPL broadband sound masking, the cognitive test scores were similar within the different talker numbers. The subjective distraction increased monotonically with the number of talkers, with higher distraction within lower SPL broadband sound masking. Overall, the results call into question the single talker assumption (being the most distracting) within the international standard for measuring open-plan office acoustic environments (ISO 3382-3:2012). Soundscapes with 4 simultaneous talkers were still not adequately providing beneficial 'babble' masking, and were more distracting than 1 active talker. In conclusion, it is suggested that the acoustics environment of open-plan offices needs better characterization by incorporating some of the complexity and psychoacoustics of multi-talker scenarios.

Crown Copyright © 2017 Published by Elsevier Ltd. All rights reserved.

## 1. Introduction

### 1.1. Background

When compared to cellular or private office configurations, there is mounting evidence against open-plan offices, in the form of demonstrable dissatisfaction amongst its occupants, and a decline in the task-based performance [1–4]. Overall, one of the main attractions of the open-plan layout – the presumed increase in productivity due to increased ease of interaction between the workers, has been shown to be substantially offset by a number of factors such as distraction due to noise and privacy issues (see reviews in [5–7]). From the perspective of a worker engaged in cognitively intensive tasks, it seems ironic that this major benefit of the shared work environment and ease of interaction, in fact, becomes the root of the problem. Intelligible speech-based communication affects both the talker and the listener (active or passive) as it has the potential to distract a *listener* who may need to concentrate on a task, and make a *talker* anxious about lack of speech privacy. This reflects in the findings of studies showing that, of all the indoor environmental quality (IEQ) factors, perceived quality of acoustics scores lowest in open-plan offices, where both intelligible speech and non-speech stimuli [7–12] have been reported as the contributing factors.

Given the significance of acoustics in the overall perception of IEQ within open-plan offices, the standardisation of some speech-related room acoustics measures in ISO 3382-3 (based largely on [13]) represents a useful development. Briefly, ISO 3382-3 considers the acoustic environment in the form of speech intelligibility and background noise distribution around workstations [14].

However, there appear to be a few fundamental issues with ISO 3382-3, arising mainly from certain simplifications and assumptions, which may limit its applicability. Specifically, the parameters in ISO 3382-3 are based around the assumption that the most distracting scenario in an open-plan office is one active talker, compared to when more than one talker is simultaneously active. This assumption seems to contradict extant literature regarding the psychoacoustics of auditory distraction in multi-talker environments (see reviews in [15,16]) such as an open-plan office. Studies have shown that a combination of few voices that are spatially seperated can be similarly distracting (in terms of subjective ratings and cognitive performance) as a single voice [17–19]. This paper explores the validity of single talker assumption within ISO 3382-3, by first presenting a literature review of multi-talker psychoacoustics, as it applies to open-plan offices. This is followed by reporting an experiment in which hypotheses that address the assumptions of ISO 3382-3 were tested within an open-plan office simulation that had a realistic multi-talker sound environment.

### 1.2. Auditory distraction due to the irrelevant sound effect

The degree of auditory distraction and the associated decline in cognitive performance within an open-plan office sound environment has been shown to be affected by the degree of uncontrolled (or, to-be-ignored) audition of irrelevant sounds: the so-called *irrelevant sound effect* (ISE; first studied by [20], see review in [21]). In this regard, what makes an irrelevant sound *stream* distracting, consistently over time, has been encapsulated within the so-called *changing-state* hypothesis (first advanced in [22]). Within the changing-state hypothesis, the sound stream can be thought in terms of segments, which vary in their acoustic-perceptual properties over time (speech being a prime example). For such segments, the extent or the degree of distraction increases with the extent to which the segments change state within the sound stream [23]. As an often cited illustration of the changing-state hypothesis, a sound stream that is composed of segments 'a b c b c a b...' has been shown to be more distractive than steady-state sound streams 'a a a a a a' (e.g., a repeating tone, or steady-state broadband noise profiles). Taking this example further, it can be argued, as was done by Macken et al. [18] (p. 45), that segments from individual voices in a multi-talker environment (each voice registering certain acoustic-perceptual attributes) can exhibit a changing-state to elicit the ISE.

It is interesting that in the presence of an irrelevant sound stream, the auditory system does not habituate to selectively limit the distraction. This is presumably due to evolutionary reasons, as a shift in relevance of the ambient sounds may sometimes require an immediate response. This applies not only in cases where there is an imminent danger, but also in more benign cases where the shift implies that the hitherto unwanted sounds begin carrying task-specific or important information. This has been typified within the duplex theory for the irrelevant sound effect. The duplex theory highlights the important differences between the *functional mechanisms* by which the irrelevant changing-state sound stream is assumed to cause distraction: either by *interference-by-process*, or *attention capture* (details in [24]). The latter relates to a momentary shift in attention from the focal task due to an unexpected change in the sound environment, which could contain either specific (e.g., their name being called) or generic cues (e.g., phone ringing) to the listeners. The former (interference-by-process) has been shown to occur when the irrelevant sounds, or segments, compete for the same cognitive processing resources as those required by the voluntary task within a certain cognitive domain, e.g., serial short-term memory [25], semantic processing [26,27]; or more typically, a combination of cognitive domains.

Since the focus in this paper is on speech-based distraction, it is worthwhile to briefly describe speech as a signal within the context of duplex theory of ISE; how its changing-state nature varies with the number of talkers; and the psychoacoustical literature relating to multi-talker environments. To begin with, speech can be considered as a 'special' signal with a high degree of redundancy in its information coding, which renders it robust against many types of system and environmental distortions, at least from the point of view of *speech intelligibility* ([15], pp. 1467). Let us consider a simple example of speech from one talker, where the energy is generally concentrated in discrete speech tokens, or segments. Such segments exhibit natural envelope modulations (or, rhythms) in the temporal and spectral domains. These spectrotemporal envelope modulations, along with the gaps between these segments typify the changing-state nature of a stream of speech, which, of course, incorporates the audition of both relevant and irrelevant speech.

Moving on from the simple case of one talker, adding more voices to the signal can have the effect of 'filling in' (or, *energetic* masking [28]) these aforementioned gaps [29]. Depending on the extent of such filling in, adding more voices can reduce the advantage (alternatively, reduce the disadvantage) of 'listening in the gaps', which, in effect, may reduce the changing-state effect. This is similar to adding spectrally shaped broadband noise, which has also been shown to be effective in reducing the 'listening in the gaps', or changing-state effect of speech [23]. Such broadband noise treatment with different kinds of spectral manipulations is generally also referred to as 'sound masking' in many commercial and scientific applications [30]. The effect of adding such 'sound masking' was also explored in this paper, as seen in Section 1.4. Another perspective in terms of adding more voices to the one talker signal can be seen as the averaging out of the modulations in individual voices, i.e., speech-on-speech *masking* over time (example of *energetic* masking), leading to reduced signal-to-noise ratio (SNR), and hence, reduced speech intelligibility. Other types of masking may include *informational* (i.e., semantic) [28] and, perhaps, *perceptual* masking ([31] p. 119), and collectively, a mixture of many voices essentially leads to the condition of speech 'babble' in the form of multi-talker speech where the streams from individual talkers may not be segmented enough to be considered distracting, and are in fact beneficial [17]. In other words, speech 'babble' has the potential to, in fact, reduce the ISE. In the context of the current experiment design, primarily the interference-by-process aspect of the ISE was more relevant, compared to the attention capture mechanism. This will be further explained in Section 2.2. However, what is important to consider here is that the conceptualisation of speech babble, thus far, has been from a largely spatial perspective, while ignoring several environmental and psychoacoustical factors; for instance, the spatial arrangement of talkers in the multi-talker babble, familiarity with voices in a real scenario that could aid segmentation, etc.

Within real listening environments, such as an open-plan office, the aforementioned speech-on-speech, or babble masking, and the purported decline in the changing-state nature of the sound stream leading to babble needs careful qualification (the following ignores any contextual issues such workplace dynamics and cultural factors, which are also likely to be quite important). One of the main reasons being that, due to the spatial separation of talkers, speech-on-speech masking is still subject to *spatial release from masking* (SRM) (see review in [31] pp. 120–124) before the sound reaches the ears (i.e., peripheral masking) [15,32]. SRM, which is essentially a monaural effect, generally means that the extent of masking decreases with a spatial separation between speech sources. The *binaural advantage* (or, disadvantage in relation to distraction), and head movements can further enhance the signal-to-noise ratio, and hence, intelligibility of this spatially unmasked speech for the listener [33].

Overall, the spatial separation of talkers and the binaural advantage can result in the restoration of, to a degree, the changing-state nature (in a sense, a 'pre-babble' state) of the combined speech signal from multiple talkers and the associated distraction due to the irrelevant sound effect. Bronkhorst noted that up to four spatially dispersed and active talkers still lead to conditions that were dissimilar to the steady-state speech-babble, albeit with a monotonic decrease in the speech reception threshold (SRT: signal-to-noise ratio required for 50% speech intelligibility) implying less benefit from any spatial release from masking, as the talker number increases [34]. However, it was also reported that there is some spread in the results of several laboratory-based studies, when they considered the effect of the number of talkers and their spatial configuration leading to masking release, with SRT increasing between 2 and 11 dB with each additional talker, which further decreased with the spatial separation of the talkers [31,32,34]. Zaglauer et al. [19] used a headphone-based open-plan office simulation to investigate the effect of voice babble on serial recall and subjective annoyance ratings [19]. Their experiment design included babble (using the same voice mixed up to six times) that was simulated from a fixed position but spatially separated from a single talker (assumed to be the most distracting, as per ISO 3382-3), which was located closer to the simulated listening position. The single talker closer to the listening position (within the distraction distance, $r_D$, as per ISO 3382-3) was simulated using the same voice as the babble voices, which, although a somewhat unrealistic scenario, represented a controlled experiment design. They found that the subjective annoyance ratings did not change with an increase in the number of voices in the babble (simulated from outside the distraction distance, $r_D$), and was not significantly different from annoyance due to a single talker. The improvement in the serial recall scores also did not show a strictly linear trend, which would be expected if the assumption of a single talker being the most distracting was true [19]. This indicates that even without spatial separation of the babble voices that leads to unmasking of individual voice, and including different voices to create a more ecologically valid babble, multi-talker scenarios nevertheless show a changing-state nature. In real office environments, hearing (or, 'glimpsing') speech segments from several spatially separated talkers within a close-range (e.g., within the distraction distance, $r_D$) may, in effect, constitute an additional factor in terms of spatial segmentation/modulations, which exhibit a changing-state nature to cause distraction, beyond the spectrotemporal modulations. There is, however, no research yet that has directly studied the changing-state nature due to spatial segmentation/modulations.

### 1.3. Speech-based distraction in ISO 3382-3: limitations and alternatives

Bearing in mind the complexity of open-plan office sound environments, and the wide range of tasks that could potentially be affected by the irrelevant sound effect (or, more appropriately here, the irrelevant speech effect), we note again that the calculations in ISO 3382-3 are specifically designed for the scenario of a single, active talker at any given point in time [14]. In terms of practicality, such an approach has many advantages, not the least of which is the fact that a single active talker simplifies the measurements and calculations involved. This is especially relevant for the metrics based on the speech transmission index (STI) in ISO 3382-3, as multi-talker STI measurement is still a relatively undeveloped concept (see Table 1 in [15], and the associated review). In particular, STI measurement does not consider the modulations of the multi-talker speech signal, which largely encapsulates its spectrotemporal changing-state nature, and the

STI standard is yet to incorporate the binaural advantage [15]. In this regard, there are alternatives and improvements to STI that characterise the speech intelligibility (or, alternatively, speech privacy) in multi-talker scenario more comprehensively (see Table 1 in [15]), and provide better prediction of the ISE [35]. However, as noted by Bronkhorst in an extensive review of this field [15], there remains no single binaural model that sufficiently addresses the complexities of multi-talker environments, including relevance of semanticity [27]. Amongst the existing models, the speech intelligibility prediction from the speech-based envelope power spectrum model (mr-sEPSM [36]) was reported by Bronkhorst [15] as being the most comprehensive model in its consideration of the complexities, while requiring only a few input parameters. The latter was used in this paper to characterise the signal-to-noise (or $SNR_{env}$ after [36]) in multi-talker environments (details in Section 2.2.3), as it considers modulations in the spectrotemporal envelope of multi-talker speech environments, which affects its changing-state nature, and consequently the irrelevant sound (or, speech) effect (ISE). The measurement method involves a single channel recording, which is simpler than a STI-based measurement and can be done in-situ with full-occupancy in offices (or similar multi-talker environments), which is currently not possible with the ISO 3382-3 method. Compared to the traditional way of understanding the ISE as the change in cognitive performance in the presence vs absence of irrelevant speech (or sound), the present paper will largely consider the ISE for single vs multi-talker speech. This is further elaborated in Sections 2.2 and 2.4.

Returning to ISO 3382-3 parameters, the main limitation is not that STI is currently inadequate for multi-talker environments, but rather the *assumption* of the single active talker scenario as the most distracting, which is inconsistent with evidence cited in Section 1.2 relating to SRM, the binaural advantage, the characterisation of speech babble, etc. Moreover, the only study cited in ISO 3382-3 ([17]; Ref. [10] therein) to justify the one-talker scenario being the most distracting, seems to be equivocal. Jones and Macken [17] addressed the role of the *number of voices* (one, two, or babble of six voices) and their *spatial location* on cognitive efficiency (i.e., SRM). Within the abstract, while it was stated that "... *with monaural presentation, 1 and 2 voices produced roughly the same degree of disruption, but babble of 6 voices reduced errors significantly*", it was also stated that "... *allocating each voice to a different spatial location restored its disruptive capacity*" (p. 216), which highlights SRM in this context, and was explored in a laboratory-based experiment 4 in the paper. While addressing practical scenarios, Jones and Macken cautioned against considering the number of voices as a "...*fixed inviolable numerical threshold above which efficiency would not decline*" (p. 224), and instead suggested that "... *a range of factors would qualify the effect of number of voices*" (p. 224) which would include voice distinctiveness, task characteristics (memory vs. reading), acoustics settings, and stationary vs. moving sources (a voice directivity issue) [17]. Furthermore, while conjecturing on the applicability of speech-on-speech, or noise-on-speech masking to decrease the ISE in office spaces, they mentioned that, "... *where conditions are such that spatial separation of the sound sources is enhanced, this advantage arising from increasing the number of voices in the signal will not be obtained*" (p. 225). In summary, Jones and Macken [17] foreshadowed the issues raised in Section 1.2 above. The findings in Zaglauer et al., as presented in Section 1.2, provided further impetus to a more systematic consideration of the distraction due to talkers in the vicinity of a listener, as is the case in most office environments [19].

### 1.4. Aims of the study

Based on the issues raised in Sections 1.2 and 1.3, this paper investigates speech-based distraction within a realistically

**Table 1**

Details of the sound environments that were simulated. The acoustic descriptor were: A-weighted equivalent SPL ($L_{A,eq}$); room noise criterion (RNC), where the parenthetical letter denoted the presence of low-frequency fluctuations (F) or not (N); and SNR in envelope-power domain ($SNR_{env}$).

| Number of active talkers (T) | Masking noise status | Sound environment label | $L_{A,eq}$ (dBA) | RNC | $SNR_{env}$ (dB) |
|---|---|---|---|---|---|
| 0 | Off | T0_M0 | 31 | 26(F) | – |
| 1 | Off | T1_M0 | 36.3 | 31(F) | 25.9 |
| 2 | Off | T2_M0 | 38.7 | 33(F) | 43.4 |
| 4 | Off | T4_M0 | 40.6 | 35(F) | 0 |
| 0 | On | T0_M1 | 42.0 | 44(N) | – |
| 1 | On | T1_M1 | 43.0 | 43(N) | 18.5 |
| 2 | On | T2_M1 | 43.6 | 41(N) | 19.4 |
| 4 | On | T4_M1 | 44.4 | 38(F) | 20.9 |

simulated open-plan office, where spatially separated talkers (potentially exhibiting SRM) were represented by calibrated loudspeaker channels. The impact of the number of simultaneous talkers, and two profiles of broadband masking noise was assessed on the cognitive performance, and subjective distraction for a group of participants. Broadband noise, which can be seen as 'filling in the gaps' that are naturally present in the spectrotemporal envelope of speech, was used to simulate masking similar to speech-on-speech masking. There is evidence, however, that such broadband masking, while effective in providing benefits against the ISE as evidenced in higher cognitive scores, was also reported as subjectively more annoying; when compared to similar maskers such as running water stream, actual speech babble, etc [37–39]. Overall, the following was hypotheses were tested:

1. *In lower SPL background noise conditions, cognitive performance would decrease with increasing number of talkers* (to the maximum of four talkers tested in this paper).

Based on the findings of studies (see Sections 1.2 and 1.3) that showed restoration of the changing-state nature of sound stream in multi-talker environments with a spatial separation of talkers, and other factors that can lead to unmasking (binaural, semantic effects, etc.), it was expected that the cognitive performance would, firstly, not improve in multi-talker environments, compared to the single-talker environments. Zaglauer et al. reported that even with low STI values, listeners were able to extract content from multi-voice babble [19]. With more than one active voice in the current experiment design, there would be a higher likelihood of the listeners tuning-in and -out of the respective 1-way conversations that were simulated (details follow in Section 2.2.1), which is consistent with the changing-state hypothesis for the multi-talker environments. Furthermore, due to the curiosity-inducing nature of such a changing-state sound stream with a high degree of semantic validity (which has been shown to affect office tasks such as proofreading [40,41]), it was expected that in some multi-talker cases, the cognitive performance, in effect, would be worse than a single-talker scenario.

2. *In higher SPL background noise conditions, cognitive performance would be similar to that observed in the single-talker or multi-talker sound environments.*

This relates to the notion of beneficial broadband masking reducing the changing-state nature, and hence, the ISE due to multiple talkers, which has been shown in previous studies [37–39].

3. *The degree of subjective distraction would increase as the number of talkers increased.*

This hypothesis mirrors the point 1 above; from the cognitive performance domain to subjective perception. Although Zaglauer et al. [19] showed similar subjective distraction due to one and many voices, the voices in their babble mix was not spatially separated, and moreover, a single speech source was used for the babble mix. It was expected that the spatial separation of voices and the use of a unique voice per talker would result in an increase in the degree of subjective distraction with an increase in the number of talkers.

4. *The degree of subjective distraction would be higher in the lower SPL background noise condition, compared to the higher SPL conditions.*

This was expected due to the reduction in the changing-state nature due to the higher SPL broadband sound masking, compared to lower SPL sound masking.

## 2. Methods

### 2.1. Laboratory simulation of an open-plan office: visual and Climactic considerations

The experiment was conducted in a climate-controlled chamber (described in [42]) that was set-up as a medium-sized open-plan office (60.6 m³; 0.7 s mid-frequency reverberation time). The climate in the chamber was maintained well within the acceptable thresholds prescribed by the relevant thermal comfort/ventilation standards [43,44] through under-floor air distribution (air temperature: 22.3 ± 0.3 °C; RH: 65.4 ± 4.5%; $CO_2$: 516.9 ± 42.2 ppm).

There were two prime objectives for the design of the chamber's fit-out – that the workstation for the participant should look and feel like a typical workstation in open office settings, and that the loudspeakers used to simulate talkers (details in Section 2.2) should not be visible to the participant at any stage. In order to achieve the first objective, the workstations were partially enclosed with medium-height partitions (1.5 m × 1.5 m each), in the form of cubicles with desks, chairs, indoor plants, and desktop computers, as seen in Fig. 1(a and b). Additionally, "skyline"® sound diffusers (RPG, Upper Marlboro, USA) were attached to the walls in order to make the sound field more diffuse and reduce any flutter echoes that normally affect rooms with parallel hard walls. The participant was seated at the location marked "P" in Fig. 1(a). There was no discernible or measureable air draft at the participant's workstation and illuminance on the desk surface was fixed at a typical office value of 500 lx. In order to achieve the second objective for the visual presentation, the participant was brought in the room through the door marked D1 in Fig. 1(a), limiting any visual access to the loudspeakers, which, along with the sufficiently high partitions, was further enforced by the participant not being allowed to move around freely in the rest of the office.

### 2.2. Laboratory-based simulation of an open-plan office: acoustical considerations

There were two distinct components in simulating the acoustics of a medium-sized open-plan office at the listening position of the participant: simulated talkers, and the spectrally shaped broadband noise for speech masking. These two components are described in Sections 2.2.1,2.2.1, respectively, leading to the creation of eight sound environments for the experiment trials, presented in Section 2.2.3.

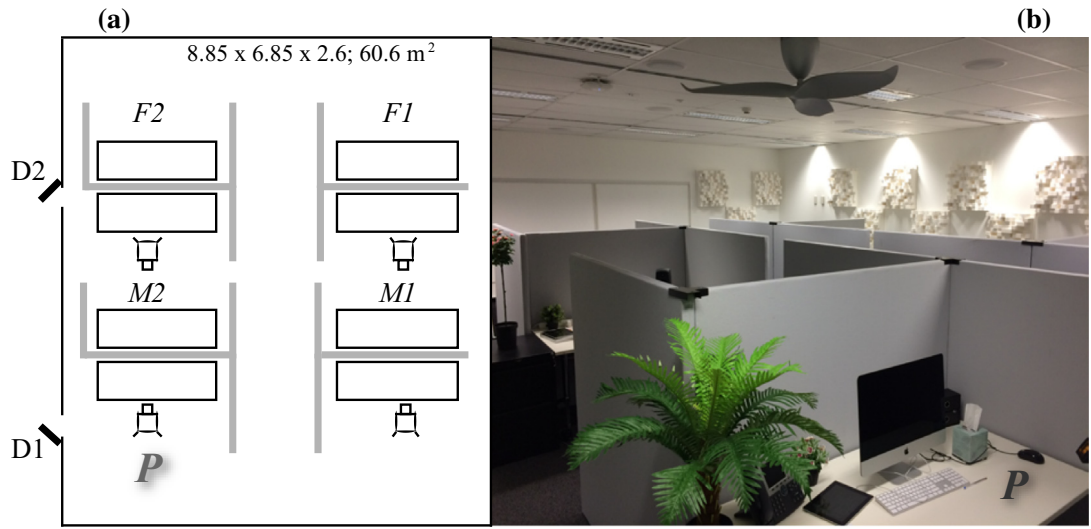

**Fig. 1.** (a) Layout and some details of the simulated open-plan office, including the representative locations of the participant (*P*), four loudspeakers that were used to simulate the talkers (two female voices: *F1*, *F2*; and two male voices: *M1*, *M2*), partitions, desks and doors (D1, D2). (b) Depicts the actual room where the participant's (*P*) desk.

### 2.2.1. Multi-talker simulation system

A 4-channel sound system was used to simulate four talkers with loudspeakers at the locations depicted in Fig. 1(a), where the 2 male and 2 female talkers that were simulated are labelled *M1*, *M2*, and *F1*, *F2* respectively. The loudspeakers were connected through a RME Fireface UFX interface (Haimhausen, Germany) to a computer running a bespoke application created using Max (Cycling '74, Walnut, USA) for signal processing. The loudspeakers used were Genelec 8010A (Iisalmi, Finland), which were placed at a height of 1.2 m from the floor, to represent seated talkers. These loudspeakers were similar to those used by Haapakangas et al. [30], and have mouth-like directivity, with dimensions comparable to a human head.

For each of these simulated talkers, the corresponding loudspeaker channel was calibrated in an anechoic room, so that its output matched the spectrum of normal speech at a distance of 1 m, as per Table 1 in ISO 3382-3 [14]. The values in this table are based on the spectrum in ANSI S3.5-1997 (R2007) [45] representing the averaged spectra of male and female voices speaking with normal effort. After the calibration, the deviations in the 125 Hz–8 kHz octave-band spectra of the loudspeaker channels were within ±1.5 dB of the spectra in Table 1 in ISO 3382-3. Essentially, calibrating each loudspeaker channel ensured that its long-term spectral output matched that of a person speaking with normal effort from that location. Note that the loudspeaker-based talkers represent a *fixed* system in space, which does not simulate the time-varying features that accompany actual human conversation (e.g., directivity, etc.). Nevertheless, the loudspeaker-based talker system facilitated a highly realistic multi-talker environment for testing participants within a factorial design.

The speech material that was used consisted of scripted two- and three-way conversations of Australian-accented English talkers (two males and females, each). These conversations were recorded over individual channels for each talker in an anechoic room (details in [46]), with omnidirectional headset microphones (DPA d:fine 66; Gydevand, Denmark) that were placed close to the talkers' mouth. The aim of recording these conversations was to simulate multi-talker speech environments that test real-world listening abilities [46]. As such, the talkers were instructed to maintain a natural mode of interaction while acting out the scripts, and rigorous quality control was maintained. The conversation scripts were deliberately quotidian, in order to maximise ecological validity. After signal processing to remove crosstalk between microphone channels, the average duration of these recordings were 3.42 ± 0.85, and 3.63 ± 0.73 min for the two-, and three-talker passages, respectively.

Sound recordings were subsequently edited to create four *configurations* of multi-talker environments (three with talkers, and one with no talkers – the *quiet* configuration); while ensuring no repetition of the speech content. Complete passages for each talker in the unedited conversations (from [46]) were manually isolated from the sound recordings, while maintaining the natural cadence, pauses, variations in speed, and interjections in these passages. Examples of complete passages (or, segments) can be seen in Fig. 2(a), which, for each talker, were one of the many isolated from their respective recordings. These passages were arranged to create 4-channel sound files (wav format, 44.1 kHz, 32 bit) of 20 min duration each, one for each talker configuration (three in total, except quiet). Only one talker from each originally recorded conversation (two- or three-way) was included in the sound file of a particular multi-talker configuration. This was done to ensure that the participant was always listening to only one part of the two- or three-way conversation recordings, which is similar to listening to one side of a telephone conversation; reported as one the most distracting speech/sound stimuli in open-plan offices ([47]; refer to [48]) for an extensive review). The four configurations of the sound environments, according to the number of talkers in their respective 4-channel sound recording used in the experiment were:

1. *0-Talker* configuration (T0): No active talkers. This *quiet* sound environment still had small fluctuations in the background noise level due to some traffic noise ingress (from one of the room's walls, which interfaced with the building façade), and the more stationary underfloor air distribution noise. A typical SPL during experiment hours at the participant's listening position was measured as 31.0 dB ($L_{A,eq}$: A-weighted equivalent sound pressure level).

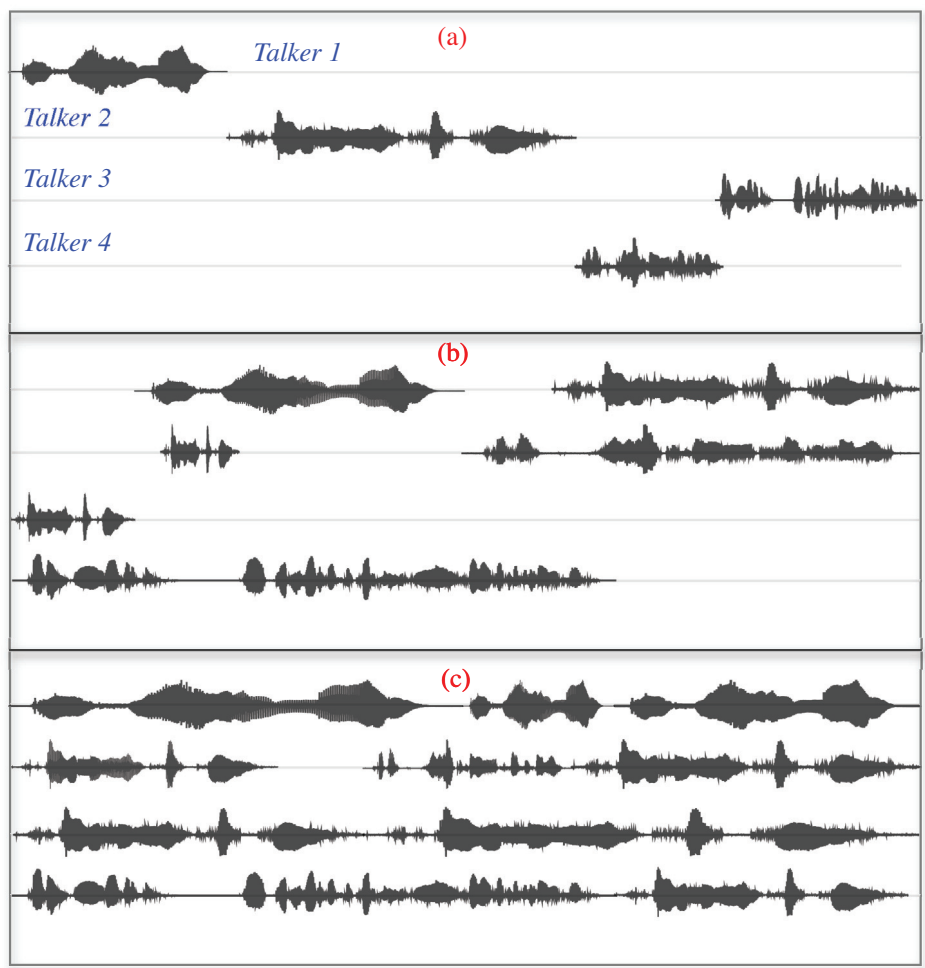

**Fig. 2.** (a) A representative arrangement of speech passages per talker (labelled Talker 1–4), where only one talker is active at any time. (b) & (c) Show two, and four simultaneously active talkers, respectively, at all times. Note that the natural pauses and interjections in the individual speech passages were maintained.

2. *1-Talker* configuration (T1): One active talker at any time, with the active talker shifting between the loudspeaker channels, depending on the length of each talking passage. The order of appearance of the talkers in the 4-channel sound file was organised in a quasi-random manner, with no overlaps in the talkers (see Fig. 2(a) for an example). Note that the 1-Talker configuration has been assumed to be the most distracting by ISO 3382-3, and several studies that cite the standard.

3. *2-Talker* configuration (T2): Same as the 1-talker configuration in all manners, except that the sound recordings were edited to ensure that 2 talkers were active simultaneously at any point in time. Fig. 2(b) shows a typical arrangement, where talkers became active in a quasi-random order, with the sole aim of maintaining two active talkers.

4. *4-Talker* configuration (T4): As seen in Fig. 2(c), 4 talkers were active simultaneously throughout, with the talker order being quasi-random. The rest was the same as the 2-Talker configuration.

### 2.2.2. Broadband noise system for providing masking from speech

To investigate the effect of broadband noise presentation (providing beneficial *speech masking*) on the performance of the participants, an overhead 13-channel sound system was used. This included 12 independently controlled flush-mounted ceiling loudspeakers (Quest QTC-2-80BC, Melbourne, Australia) and one channel that controlled 4 custom-built subwoofers (installed close to each other), which were hidden above the suspended ceiling in the simulated office. The presence of masking noise was not explained as part of the experimental design to the participants. Instead, due to the unobtrusive placement of the loudspeakers, and the actual underfloor air-conditioning system being relatively quiet and steady (31 dBA), it was assumed that the noise would be attributed to typical overhead HVAC operation by the participants. In this regard, Max/MSP was used to generate spectrally balanced noise (approximately −5 dB per octave slope, as per [30]) from these loudspeaker channels (incoherent between channels), which was recorded during daytime at the participant position as 42 dBA, which included noise from the underfloor air distribution system and any external traffic noise.

The increase in the voice level of the talkers in the presence of increased background noise levels, known as the Lombard effect (see [49] for a review), was not implemented for the simulated talkers in this experiment.

### 2.2.3. Sound environments for the experiment trials

As seen in Sections 2.2.1 and 2.2.2, there were four talker configurations, which were labelled T0, T1, T2 and T4, based on the number of active talkers; and two masking noise configurations, which were labelled M0 and M1 (off, or on, respectively). This led to eight sound environments in a factorial format, which constituted the eight experiment trials (details in Table 1). Besides $L_{A,eq}$, the other acoustic parameters used to characterise these environments, all derived from long recordings (up to 5 min) at the listening position of participant (1.2 m from the floor; representing a seated height), included the following:

1. *Room noise criterion* (RNC), which is a method of noise assessment based on the more common Noise Criterion (NC) method, but with a psychoacoustically more precise approach, especially in detecting low-frequency fluctuations NC, NCB, etc.) [50]. The parenthetical letter represents whether or not the RNC algorithm detected substantial random fluctuations (F) or surging within the low-frequency octave bands (N: for neutral). Note that these low-frequency fluctuations cannot be represented within A-weighting.

2. Since an STI-based method for multi-talker environments is not yet viable, a more valid objective method: the speech-based multi-resolution envelope-power based model (mr-sEPSM [36]), was used. While the aim of mr-sEPSM is predicting speech intelligibility (SI) [51], it remains informative in the current scenario where the effect opposite to SI (i.e., speech privacy) would be desirable (in the open-plan office environments).

Briefly, the model requires signals representing *noise*, and *speech + noise*, to output the SNR in the envelope-power domain, or $SNR_{env}$, following stages of peripheral filtering (to represent basilar membrane processing), envelope extraction, and further filtering (specific to speech audition) in modulation bands (1–64 Hz; octave-spaced) and carrier-frequency bands (63–8000 Hz center-frequencies; 1/3 octave-spaced). The key feature of this approach is that the model accounts for time-varying auditory effects of both *noise* and *speech*, within the envelope and frequency bands. Here, the *noise* signal could include irrelevant speech (if so designed), and the *speech* signal could include speech from multiple sources, which makes the mr-sEPSM approach especially attractive for open-plan office environments. In comparison, STI currently only accounts for fluctuations in frequency bands (not within the time-domain signal), and only for *speech*. $SNR_{env}$ also accounts for room reverberation (as does STI), spectral subtraction (STI does not), speech-on-speech unmasking (STI does not). While it was not presented in the current paper, the $SNR_{env}$ can be converted, using simple parameters, into percentage of correctly recognised speech for a statistically 'ideal observer' (which implicitly models the psychometric function), while accounting for the performance increase due to the reduction of the number of response alternatives.

In the current paper, the *noise* input to the mr-EPSM model was recordings T0_M0 and T0_M1 (i.e., no speech), which were compared with corresponding *speech + noise* recordings (i.e., $SNR_{env}$ derived for T0_M0 compared with T[1 or 2 or 4]_M0; and T0_M1 compared to T[1 or 2 or 4]_M1). These $SNR_{env}$ values (in dB) are listed in Table 1.

Note that the sound recorded at the listener's position for the different number of talkers (T1, T2, T4) was not equalised to have the same SPL. This was done in order to maintain ecological validity, as increased sound levels would be expected when there are more active talkers in the sound environment (but the Lombard effect was not incorporated). A similar approach was taken in [39]. Moreover, numerous studies have shown that the SPL changes within a reasonable range do not affect the ISE ([29,52]; review in [53]). The combined SPL (masking noise and multi-talker speech) was also maintained close to, but under ~45 dBA (Table 1), which has been reported as sufficiently effective for masking, while not being perceived as 'annoying' by the participants [54,55].

## 2.3. Participants

Thirty participants (15 Female; Age range 18–39 years, mean: 23 years, and standard deviation: 6 years) volunteered for the experiment[1]. None of the participants reported any hearing impairment. The participants were native English speakers (with an Australian accent); students in fields of Architecture, Design and Planning; had some experience of working in open-plan offices or studios; and were paid for their time. The participants were asked to adhere to a formal dress code, typical for office work during the late summer/early autumn period in Sydney, Australia, corresponded to an insulation value (clo) of approximately 0.6, plus another 0.1 clo for the insulation of the chair on which they were seated during the experiments. Only one participant was tested at a time, with the entire duration of the experiment being four hours per participant including orientation, practice of the experiment tasks, and structured breaks.

## 2.4. Experimental tasks

For the eight experiment trails (simulated sound environments in Table 1), the subjective assessments and cognitive tests performed by the participants were as follows.

### 2.4.1. Subjective assessments

1. **Noise questionnaire**: This has been used in previous studies as a reliable method to determine the noise sensitivity of a person overall; and also within the domains of leisure, work, habitation, communication, and sleep [10,30]. The noise questionnaire, which was translated from German (NoiSeQ; [56]), included thirty-five items, each rated on a seven-point scale (*Completely Disagree* through to *Completely agree*). This questionnaire was administered before the experiment trails in quiet ambiance (T0_M0 in Table 1).

2. NASA Task Load Index (**NASA-TLX**): This questionnaire was administered after each experiment trail (i.e., eight times in total) on a bespoke iPad application, was used for a subjective assessment of the *workload* that the participants experienced after performing the cognitive tasks within a particular sound environment, (represented by each experiment trial). The meaning of workload, and how it was encapsulated in the NASA-TLX was explained to the participants, as described in [57]. Briefly, NASA-TLX involves the participants providing an assessment of their perceived workload for a particular task, measured across six dimensions (mental, physical and temporal demand, performance, effort and frustration). The participants assigned relative weightings to the six dimensions.

3. **Sound Environment questionnaire**: The participants filled out a sound environment questionnaire on a bespoke iPad application after each experimental trial (eight times in total). The perceived disturbance due to each sound environment was rated with four items, each presented on a seven-point scale, based on the findings of Haapakangas et al. [30] who reported them as the most reliable out of several items that were rated. The items were: 'The sound environment was pleasant'; 'The sound environment impeded my ability to concentrate'; 'It was easy to get used to the sound environment over time'; 'My attention was drawn towards the sound environment'.

### 2.4.2. Cognitive tests

While typical office workers may perform diverse activities, varying in their cognitive load, studying the ISE on each of these activities was not feasible. A more pragmatic approach involved studying the variation of a range of cognitive skills generally required in office tasks. Hence, the cognitive test chosen encompassed language comprehension, active learning, problem solving, reasoning, planning, etc. The tests were administered through the Cambridge Brain Science (CBS; www.cambridgebrainsciences.com) website, with each test representing an interactive algorithmic implementation of a certain cognitive psychology paradigm. In total, the participants performed seven short tests within each trial (each trial being a simulated sound environment), which quantified their cognitive problem solving skills within generic categories of *memory, reasoning, concentration,* and *planning*. Within each of these categories (as used by CBS), the selected tests varied in their durations, and are briefly described as follows (details in [58]):

1. Memory: The *Digit Span* ($DS_M$) test involved a *forward recall* task, which assessed short-term information storage. The design of this test differs slightly from the serial recall tests that have previously been used in studies of ISE (summary in [48]). For each recall problem, the participants recalled a certain number of digits that were presented on a computer screen in the order of appearance. If the participants recalled the sequence correctly, the number of digits in the subsequent problem increased by one; otherwise it decreased by one. For this test, there was no time limit on the recall, but the test finished when the participant failed for a total of three times in recalling the correct sequence of digits. The final score was the number of digits that the participant successfully recalled. This differs from the design of serial recall tests that have previously been used in studies of ISE, which generally include a time bound recall of a fixed number of digits (summary in [48]).
2. Reasoning: Two tests were selected that incorporated decision-making, based on deductive reasoning for visual information. The tests are designed to invoke verbal processing for reasoning, which can be assumed to be important within the ISE framework. The *Grammatical Reasoning* ($GR_R$) test involved the participant participants choosing whether a statement involving a visual comparison was 'true' or 'false'. The other test in this category was called *Double Trouble* ($DT_R$), where the participants had to select the correct ink colour of a word that was presented, out of two words that were presented.
3. Concentration: Two tests were selected that assessed the participants' skills relating to visuospatial perception and mental processing of complex objects. In the *Rotations* ($Rot_C$) test, the task was to assess whether one of the two boxes filled with varying number of coloured squares, if rotated in space, could match the other box. The cognitive abilities required in this test have been shown to be highly correlated with tasks such as 'route planning' [59]. The other test in this category was called *Feature Match* ($FM_C$), where the participants were presented with two separate boxes, each with an array of abstract shapes, varying in complexity for each problem. The task was to ascertain whether the two boxes were identical.
4. Planning: Two tests were selected that assessed the participants' skills in complex decision making that is generally involved in *planning ahead* to reach a certain goal, using the least number of tries. The *Hampshire Tree Task* ($HT_P$) required the participants to use least possible number of tries to rearrange numbered balls presented in a jumbled format, to match a predetermined sequence. The *Spatial Slider* ($SS_P$) test involved the participant being presented a certain arrangement of a grid of numbers for a short duration, which was then jumbled. The task then required the participants to rearrange the jumbled grid to its original form, in the least possible tries.

### 2.5. Procedure

The experiments were conducted in two blocks: one morning and one afternoon *session*; each being four hours in duration. Each of these sessions was attended by a single participant (i.e., four hours per participant, over either the morning or the afternoon session). Before the experiment trial, the participants filled out the *noise questionnaire* (item 1; Section 2.4.1), and were given an orientation of all the task, as detailed in Section 2.4. This was followed by a practice run in which the participants performed all the experiment tasks in a certain sequence, within the T0_M0 sound environment; the quietest sound environment. Note that it was not assumed here that the quietest sound environment represented the optimum ambiance for cognitive tasks. In the practice run, the participants familiarised themselves with all the cognitive tests; to the point where their scores were stable over retests to minimise any learning effect. As depicted in Fig. 3, each simulated sound environment comprised of a certain masking noise state. The masking noise was *off* for the first half of each session to enable a smooth transition from practice (done with no masking noise) to trials. The second half of each session was preceded by a short break, during which the participant left the simulation to an adjacent room. The experimenter turned the masking noise *on* for the second half of each session, while the participant was not present in the simulation. After their break, the participant returned to the simulation that had the masking noise turned *on*, which was considered to be a smoother transition than the alternative of having an abrupt change in the state of masking noise (from *off* to *on*) with the participant present in the simulation. As mentioned in Section 2.2.2, the reproduction of the masking noise was designed so that the participant were most likely to attribute it to HVAC operation. Each masking noise state (*on* and *off*) had a certain number of simulated talkers (0, 1, 2, or 4), where the presentation order of the number of simulated talkers was randomised.

Within the simulated sound environments, the participants performed the sequence of tasks depicted in the box with the dotted boundaries in Fig. 3. They consisted of the seven cognitive test (order randomised; each test performed once per trial) performed online on an iMac computer, followed by subjective assessments that included the NASA-TLX questionnaire and the sound environment questionnaire, both presented on an iPad. The duration of each simulated sound environment was 20 min, which included approximately 15 min for performing the cognitive tests, and 5 min for filling the NASA-TLX and sound environment questionnaire (items 2 and 3, respectively; Section 2.4.1).

After the eight experiment trials, the participants filled out the IEQ questionnaire (item 4; Section 2.4.1) indicating the degree of distraction from each of the simulated talkers.

### 2.6. Data analysis

Due to the hierarchical and repeated-measurement nature of the experiment (leading to autocorrelation), linear mixed-effects models with varying complexities, in terms of the hierarchy of *fixed* and *random* effects, were fit to the data. In the context of the current experiment, mixed-effects models allowed estimation of the average intercepts and slopes of the independent variables as *fixed-effects*, while the *random-effects* accounted for the deviation of each participant from the average intercept and slope. Note that the modelling stages involve first fitting a general linear model (GLM) to the data, and comparing it to a mixed-effects model, and only using the latter if it represents a better model than the former. In this regard, a statistical model without the random-effects depicts the classical linear regression model (i.e., GLM), whereas modelling the random-effects accounts for the natural heterogeneity in the participants across the following independent variables (which are also assigned labels, as seen below) in the experiment:

- *Gender.* 2 × groups (M and F).
- *NoiseSens.* 2 × groups, representing the noise sensitive (NS1) and non-sensitive (NS0) groups. The group allocation was based on a median split of the participants' cumulative rating for the noise questionnaire items.
- *Session.* 2 × groups, representing whether the participant attended the morning (SesM; 0900–1300 h.), or the afternoon (SesA; 1300–1700 h.) session.
- *MaskingNoise.* 2 × groups, representing whether the overhead masking noise was on (M1) or off (M0).
- *NumTalk.* 4 × groups, representing the number of talkers that were simulated in the sound environment.
- *SoundEnv.* 8 × groups, which resulted from combining the respective groups of the *MaskingNoise* and *NumTalk* variables, as seen above.

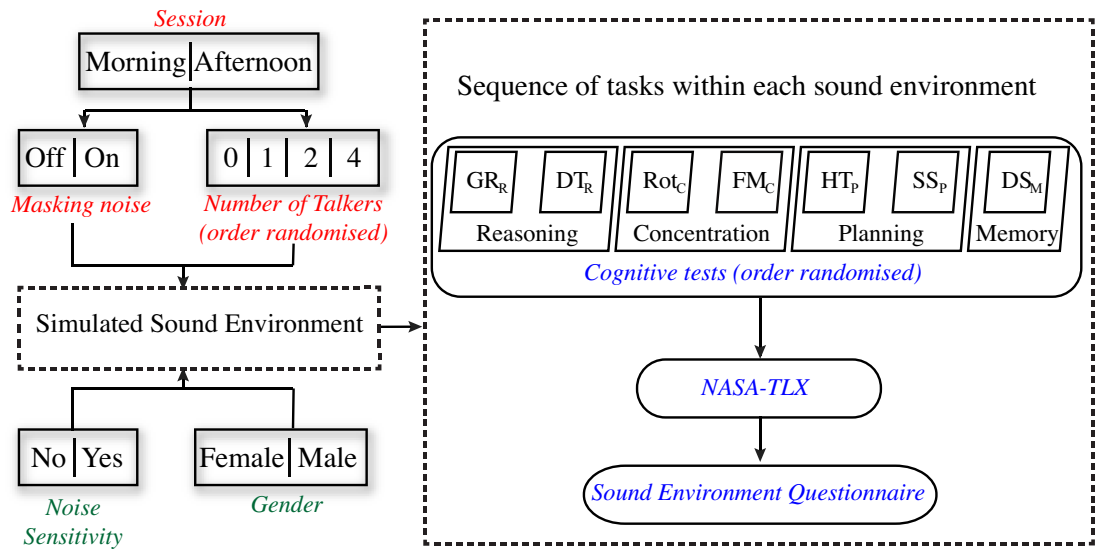

**Fig. 3.** The experiment plan. Each simulated sound environment (broken rectangle on the left) is depicted as varying due to the interaction of the independent variables (*Session, Gender*, etc.). The sequence of tasks within each sound environment is depicted within the broken rectangle on the right.

All the analyses were done within the R software, using the function *lme()* for the mixed-effects modelling, which is included in the *nlme* package [60], with the *maximum likelihood* method for estimating the model parameters in the analysis. The *ggplot2* package [61] was used for plotting and exploring relationships between the variables, and the *dplyr* package [62] was used for data management.

Two sets of statistical analyses were performed, each set modelling a different dependent variable, as follows:

1. *Cognitive test scores:* Here, the scores of each cognitive test was used as the dependent variable, representing an objective variation in the cognitive performance due to variation in the sound environments. As such, the performance of separate cognitive abilities (represented by the corresponding CBS tests) was modelled separately, e.g., the scores of the grammatical reasoning test (GR$_R$; Fig. 3) modelled as a function of the independent variables.
2. *Subjective auditory distraction:* Here, the subjective auditory distraction, calculated as the mean rating of the items of the *sound*

*environment questionnaire* (item 3, Section 2.4.1; and Fig. 3) was used as the dependent variable. This questionnaire, as mentioned in Sections 2.4.1 and 2.5, was answered once per sound environment, after all the cognitive tests had been performed, and can be seen to represent the participants' overall subjective auditory distraction due to the sound environment.

Each statistical analysis started with a model with just the intercept, followed by introducing additional independent variables, both as fixed-effects, and random-effects within a nested hierarchy. The performance of the models after introducing each independent variable (fixed-effects, their interactions, and random-effects) was compared using the Akaike's Information Criterion (AIC) [63]. AIC is a measure that allows comparing the goodness-of-fit of competing models against the number of variables used per model (in other words, the models' complexity, or degrees of freedom). Simply, if introducing an independent variable in a model did not statistically reduce the AIC measure from the previous model (without the independent variable), this additional variable was not included in the model, and vice versa. Orthogonal contrasts were designed for the variables with more than two levels, to test the hypotheses of this paper (Section 1.4). In addition, the *workload* calculated for each sound environment, based on the NASA-TLX ratings, was used as a covariate in the analyses.

## 3. Results and discussion

### 3.1. Mixed-effects models

In the following, due to the large number of fixed-effects and their interactions, the results for only the highest-order interactions are presented, which always include the lower-order interactions. The focus is on discussing the results in terms of the expectation presented in Section 1.4, which was deemed more practical here than a detailed account of the mixed-modelling steps (which were briefly described in Section 2.6).

#### 3.1.1. Cognitive test scores

The scores for the grammatical reasoning (GR$_R$) cognitive test, as the response variable, showed a significant variation in its intercepts for the random-effects, which were modelled as a nesting of

**Table 2**

Contrasts of the fixed-effects groups and their interactions; showing the $b$ coefficients, standard errors (SE) of $b$, $t$-statistics (degrees of freedom in parentheses), $p$-values (2-tailed; significance at $p < 0.05$ highlighted in a bold font), and effect size ($r$; where $r = 0.5$ shows a *large* effect accounting for 25% of the total variance, $r = 0.3$ shows a *medium* effect accounting for 9% of the total variance, and $r = 0.1$ shows a *small* effect accounting for 1% of the total variance [64]).

| | $b$ | $SE(b)$ | $t(175)$ | $p$ | $r$ |
|---|---|---|---|---|---|
| (A) $GR_R \sim SoundEnv * NoiseSens * Gender$ | | | | | |
| 1. (T1_M0 vs. T2_M0) × (NS0 vs. NS1) × (F vs. M) | 3.9 | 2.0 | 1.9 | **<0.05** | 0.1 |
| 2. (T1_M0 vs. T4_M0) × (NS0 vs. NS1) × (F vs. M) | 7.3 | 2.0 | 3.6 | **<0.01** | 0.3 |
| 3. (T2_M0 vs. T4_M0) × (NS0 vs. NS1) × (F vs. M) | 5.6 | 1.8 | 3.2 | **<0.01** | 0.2 |
| 4. (T1_M1 vs. T2_M1) × (NS0 vs. NS1) × (F vs. M) | 0.9 | 2.0 | 0.4 | 0.67 | |
| 5. (T1_M1 vs. T4_M1) × (NS0 vs. NS1) × (F vs. M) | 1.4 | 2.0 | 0.7 | 0.48 | |
| 6. (T2_M1 vs. T4_M1) × (NS0 vs. NS1) × (F vs. M) | 0.3 | 1.8 | 0.2 | 0.87 | |
| | $b$ | $SE(b)$ | $t(189)$ | $p$ | $r$ |
| (B) $Rot_C \sim SoundEnv * NoiseSens$ | | | | | |
| 7. (T1_M0 vs.T2_M0) × (NS0 vs. NS1) | 20.1 | 10.7 | 1.9 | **<0.05** | 0.1 |
| 8. (T1_M0 vs. T4_M0) × (NS0 vs. NS1) | 20.7 | 10.7 | 1.9 | **<0.05** | 0.1 |
| 9. (T2_M0 vs. T4_M0) × (NS0 vs. NS1) | 20.4 | 9.3 | 2.2 | **<0.05** | 0.1 |
| 10. (T1_M1 vs. T2_M1) × (NS0 vs. NS1) | 18.9 | 10.7 | 1.8 | **<0.05** | 0.1 |
| 11. (T1_M1 vs.T4_M1) × (NS0 vs. NS1) | −13.7 | 10.7 | −1.2 | 0.2 | |
| 12. (T2_M1 vs. T4_M1) × (NS0 vs. NS1) | −16.3 | 9.3 | −1.7 | 0.08 | |

Session within SoundEnv, within NoiseSens, within Participants ($\chi^2$ (6) = 136.4, $p < 0.0001$).

For the fixed-effects, Table 2A shows the independent variables in the final model along with their interactions that reached significance ($\chi^2$ (37) = 85.5, $p < 0.0001$), with no other variable or interactions reaching significance, or improving the AIC. For the Rotations ($Rot_C$) cognitive test as the response variable, the test scores showed a significant variation in its intercepts for the random-effects, which were modelled as a nesting of *Session* within *SoundEnv*, within *NoiseSens*, within *Participants* ($\chi^2$ (6) = 52.9, $p < 0.0001$). For the fixed-effects, Table 2B shows the independent variables in the final model along with their interactions that reached significance ($\chi^2$ (21) = 24.1, $p < 0.05$), with no other variable or interactions reaching significance, or improving the AIC. Note that, for brevity, $\chi^2$ statistics for the inclusion of each variable as fixed-effects in the model, and their interactions, is not provided here. Contrasts were used to compare the fixed-effects parameter groups, which are depicted in Table 2. Although there were significant lower-order interactions and fixed-effects variations, they are not interpreted here due to the presence of higher-order interactions that included them, which were:

1. Interaction 1 and 2 in Table 2 shows that the $GR_R$ scores were significantly lower for the sound environment T2_M0 (two active talkers with masking noise off; *small-sized* effect) and T4_M0 (*medium-sized* effect) compared to T1_M0, for noise-sensitive (NS1) compared to noise-insensitive (NS0) participants, and for male (M) participants compared to females (F).
2. Interaction 3 shows that $GR_R$ scores were significantly lower (a *medium-sized* effect) for T4_M0 compared to T2_M0, for noise-sensitive (NS1) compared to noise-insensitive (NS0) participants, and for male (M) participants compared to females (F).
3. Interactions 4–6 show that $GR_R$ scores for the T2_M1, T4_M1 sound environments were not significantly different from T1_M1.
4. Interaction 7 and 8 show that the $Rot_C$ scores were significantly lower for T2_M0 and T4_M0 (both *small-sized* effects) compared to T1_M0, and for noise-sensitive (NS1) compared to noise-insensitive (NS0) participants.
5. Interaction 9 shows that the $Rot_C$ scores were significantly lower (a *small-sized* effect) for T4_M0 compared to T2_M0, and for noise-sensitive (NS1) compared to noise-insensitive (NS0) participants.
6. Interaction 10 shows that the $Rot_C$ scores were significantly lower (a *small-sized* effect) for T2_M1 compared to T1_M1,

and for noise-sensitive (NS1) compared to noise-insensitive (NS0) participants.
7. Interactions 11, 12 show that $Rot_C$ scores for T1_M1, T2_M1 sound environments were not significantly different than T4_M1.

Collectively, points 1,2,4,5 suggest that, with the masking noise turned off, there was a significant decline in the cognitive test scores from the one-talker environment to the two- and four-talker sound environments. This is in direct disagreement with the assumption in ISO 3382-3 listing the one-talker scenario as the most distracting (as far as decline in cognitive performance is concerned). Not only did the cognitive performance not increase with the increase in the number of talkers, it showed the opposite trend, which has not been shown in previous studies, but was mentioned as a likely scenario in hypothesis 1 in Section 1.4. While the interpretations of the interactions in Table 2 may seem somewhat subtle (characteristic of interpreting higher-order interactions), overall, the results support hypothesis 1 in Section 1.4 within two cognitive tasks ($GR_R$ and $Rot_C$). The significant results for the $GR_R$ and $Rot_C$ tasks, which were listed under the generic categories of *reasoning* and *concentration*, respectively, point towards the interference-by-process affecting these tasks; though this needs to be explored further.

Spatial release from masking (or, SRM) is expected in the multi-talker sound environments, due to the sparse talker placement in the simulated office space. Since SRM is incorporated within mr-EPSM, it is not straightforward to explain why T4_M0 had the worst cognitive performance, despite having relatively poorer $SNR_{env}$ (Table 1). This, while highlighting limitations currently in the mr-EPSM approach to characterise multi-talker environments, also suggests the contribution from other factors. In this regard, future inquiry is suggested for the relative contribution of unmasking from various factors such as SRM, semantic context of the ISE, perceptual masking/unmasking, etc.; and indeed other environmental factors (acoustic, or otherwise) not considered in this paper. There may also be a need for soundfield analyses that focus on quantifying the changing-state nature of distracting speech from spatially spread multiple sources, similar to [65] and [66].

Another area of research would be exploring the threshold beyond which the changing-state (or segmented) nature disappears towards a uniform *babble* that can mask irrelevant speech. The relevant dimension for such a threshold may include the number or talkers, their spatial arrangement, spectrum, and relative or overall levels. Although Zaglauer et al. [19] had explored the effect on cognitive performance due to an increase in the number of voices that constituted babble, the babble voices in their study were not spatially separated. Moreover, the babble was created by mixing the same voice to increase the number of voices. The current results show that the spatial location and a semantically valid presentation of voices needs to be considered when exploring the babble effect. Such an approach, while representing a more intensive experimental design for a laboratory based study, is obviously closer to the natural occurrence of voices and semantic contexts in real office environments.

With the noise masking turned on, points 3 and 7 indicate that the scores of the cognitive task were not significantly different across the various talker configurations. This is consistent with previous research, where similar broadband noise sources have been shown to be effective in masking speech from one or multiple sources, and leading, somewhat, to improved cognitive performance [30][38]. However, point 6 indicates that there was a significant decline in the cognitive scores from the one-talker to two-talker configurations. This suggests that, even with additional broadband masking, the effect of a multi-talker sound environment should not be ignored. It is likely that speech from two talkers may still be segmented enough to cause performance decline, though this number of talkers is not suggested as a threshold here, beyond which noise masking is effective (also see Section 3.1.2, with respect to point 3 in Section 1.4). However, apart from point 6, the results within two cognitive tasks (GR_R and Rot_c) are in agreement with hypothesis 2 presented in Section 1.4, where masking noise was posited (based on extant research) to remove the detrimental effect of multiple talkers on cognitive performances. Note that while broadband noise maskers (such as pink noise) have been shown to have positive effect on cognitive performance, they are generally not reported as subjectively pleasant, or increasing the preference of the resulting sound environment [37–39]. This highlights the need to consider both task performances and subjective opinions (see 3.1.2) in characterising a complex sound environment, such as an open-plan office.

For the other cognitive tests in this experiment (besides GR_R and Rot_c), the analyses showed a more-or-less similar trend, with multi-talker environments generally showing lower scores than one-talker environment, with the disparity diminishing with additional masking noise. These trends, however, were not statistically significant at 95% confidence. Most notably, the DS_M test, which is a modified version of the serial recall task that has consistently been shown to be affected by the ISE in previous studies (Section 2.4.2), did not reach statistical significance. This is attributed here to the difference in the designs of the DS_M and the more classic serial recall task.

### 3.1.2. Subjective auditory distraction

The subjective auditory distraction ratings showed a significant variation in its intercepts for the random-effects, which were modelled as a nesting of *Session* within *SoundEnv*, within *NoiseSens*, within *Participants* ($\chi^2$ (6) = 50.17, $p < 0.0001$). For the fixed-effects, Table 3 shows the final model that reached significance ($\chi^2$ (13) = 41.1, $p < 0.0001$), with no other variable or interactions reaching significance, or improving the AIC. Contrasts were used to compare the fixed-effects parameter groups, of which the contrasts that reached significance are depicted in Table 3.

It must be noted here that the auditory distraction scores used in the analysis above represented subjective ratings for the entire duration of the simulated sound environment (approximately 20 min). Hence, these ratings encapsulated an overall sense of subjective distraction that the participants experienced while doing the cognitive tasks within a particular sound environment. Contrasts 2 and 3 show that four simultaneously active talkers were significantly more distracting than the one talker environments, for both the masking noise off (T4_M0; *small-sized* effect) and on (T4_M1; *medium-sized* effect) sound environments, which is also depicted in Fig. 4. These findings partially support hypothesis 3 (Section 1.4) in that the subjective distraction increases with the number of talkers, as the talker number increases from one to four, but not from one to two, or two to four. This is interesting, as an increase in subjective distraction from one to four talkers here is consistent with the lowest cognitive scores for T4_M0 (Table 2), which was not explained directly with the SNR_env being the lowest (value of 0) for this case. Furthermore, it reinforces previous findings that both subjective and objective scores are critical in studying the effects of open-plan office sound environments, as while the cognitive scores underscore performance decline for specific tasks, they are, at best, only presenting an incomplete view of the complex psychophysical state of a participant, which is better represented though subjective ratings [67].

The effect of spatially separating the talkers, and the more ecologically valid experiment design may account for why the subjective ratings in Zaglauer et al. [19] did not vary with an increase in the number of voices in the babble speech, but did show a decline in the current study. However, even with similar distraction ratings for scenarios with one voice compared to multi-voice babble, they suggested further research that address reduction of distraction (through babble masking, etc.) and improving worker comfort with more realistic experiment designs [19]. This is also the main recommendation of the current study, albeit the current findings suggest a more refined focus on the psychoacoustics of multi-talker environments.

As per contrast 1, support for hypothesis 4 (Section 1.4) was more straightforward, with the sound environments with lower level of background sound masking that do not sufficiently mask the irrelevant speech (from one, or multiple talkers) being reported as subjectively more distracting. The support for hypotheses 3 and 4 (Section 1.4) here that addressed subjective assessments, combined with the support for points 1 and 2 that addressed cognitive performance, further call into question the single-talker assumption of ISO 3382-3.

### 3.2. Limitations and general discussion

Similar to some previous studies ([16][30][68], the variable *workload* resulting from the NASA-TLX questionnaire did not contribute as a variable in any of the statistical models, suggesting limited applicability in assessing open-plan office environments. However, it may be possible to adapt the NASA-TLX style questionnaire, by substituting (perhaps) certain factors such as 'physical effort' that may not be that critical in office settings.

**Table 3**
Contrasts of the fixed-effects groups and their interactions; showing the *b* coefficients, standard errors (SE) of *b*, *t*-statistics (degrees of freedom in parentheses), *p*-values (2-tailed), and effect size (*r*; where *r* = 0.5 shows a *large* effect accounting for 25% of the total variance, *r* = 0.3 shows a *medium* effect accounting for 9% of the total variance, and *r* = 0.1 shows a *small* effect accounting for 1% of the total variance [64]).

| Auditory distraction ~ SoundEnv | *b* | *SE(b)* | *t*(196) | *p* | *r* |
|---|---|---|---|---|---|
| 1. M0 vs. M1 | 0.2 | 0.1 | 3.2 | < 0.01 | 0.2 |
| 2. T1_M0 vs. T4_M0 | −0.2 | 0.1 | −1.7 | < 0.05 | 0.1 |
| 3. T1_M1 vs. T4_M1 | −0.3 | 0.1 | −2.5 | < 0.01 | 0.2 |

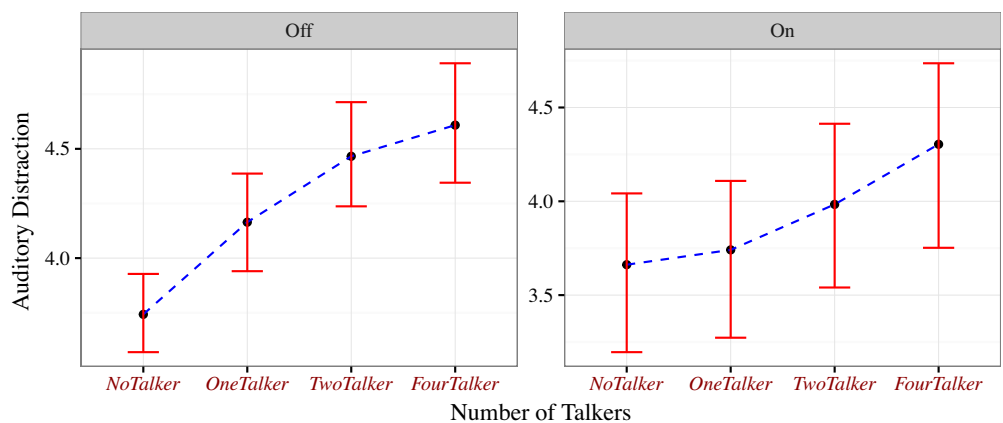

**Fig. 4.** The left panel shows the subjective *auditory distraction* ratings plotted for the different *number of talkers* while the masking noise was turned 'off', and the right panel showing the corresponding ratings with the masking noise turned 'on'. The points on the graphs show the mean values, with the error-bars depicting the 95% confidence interval (calculated using bootstrapping).

Since the aim of the experiment design was to achieve a high degree of realism, it was not possible to partial out the relative contribution of several factors that may have contributed to the cognitive scores and subjective ratings. Such factors, including the degree of spatial separation of talkers, the effect of semanticity, the threshold for the babble effect, etc., are suggested as experiment variables in future studies.

The current open-plan office simulation was constrained by the physical size of the laboratory room, and, as such, the context would be limited to small to medium sized office floors. No generalisation to include larger office floors is intended here, which might be subject to further considerations. Due to time limitations, the experiment design did not feature other important office soundscape elements that could contribute towards the ISE, such as telephones ringing, or in general, the attention capture mechanism though which ISE operates. This could be included in future implementations, along with cognitive tasks such as writing (as in [69]) and proof-reading that could be affected more strongly due to the interference-by-process aspects of the ISE.

## 4. Conclusions

The aim of the paper was to compare, within a realistic open-plan office simulation, the effect of having more than one talker in the sound environment, on both the cognitive performance and auditory distraction perceived by participants.

The results, in particular the subjective auditory assessments by participants, showed that multi-talker environments were perceived as more distracting, and had lower cognitive performance, when compared to the one-talker environments. This is pointing towards reconsideration of crucial assumptions within ISO 3382-3 [14], and research conducted within the open-plan office paradigm, where the psychoacoustics of multi-talker environments has been (incorrectly, given the current results) more-or-less ignored, or oversimplified. Speech intelligibility (and privacy) within multi-talker environments is an inherently complicated issue, affected by factors such as the spatial location of talkers, the spectrotemporal modulations, etc. However, comprehensive binaural models such as those suggested by [15] may, in the long run, be better suited to study and characterise the irrelevant sound effect in multi-talker environments, especially due to the fact that they can be run in-situ with or without the office occupants. This would be subject, of course, to field testing in real open-plan office environments.

## Acknowledgments

This research was supported by the Australian Research Council's Discovery Projects funding scheme (project DP160103978).